\documentclass[article]{agujournal2019_arxiv}
\usepackage{url} 
\usepackage{lineno}
\usepackage{soul}

\usepackage{graphicx}
\usepackage{amsmath}
\usepackage{mathtools}

\usepackage{natbib}
\newcommand{\vect}[1]{\boldsymbol{#1}}
\usepackage{mathtools,bbm}

\usepackage{amssymb}
\usepackage{gensymb}

\usepackage[nomarkers,figuresonly]{endfloat}

\usepackage{amsmath}

\usepackage{bbm}

\usepackage{float,graphicx}
\usepackage{tabularx, booktabs}

\usepackage{xcolor}

\usepackage{algorithm}
\usepackage[noend]{algpseudocode}

\usepackage{footnote}

\algrenewcommand\algorithmiccomment[1]{\hfill \(\phantom{ }\) #1}

\newcommand{\algorithmfootnote}[2][\footnotesize]{%
  \let\old@algocf@finish\@algocf@finish
  \def\@algocf@finish{\old@algocf@finish
    \leavevmode\rlap{\begin{minipage}{\linewidth}
    #1#2
    \end{minipage}}%
  }%
}

\draftfalse

\journalname{JGR: Machine Learning and Computation}

\usepackage{hyperref}

\hypersetup{
    colorlinks,
    citecolor=black,
    linkcolor=black
}

\begin{document}

\title{Deep Learning Forecasts Caldera \\ Collapse Events at Kīlauea Volcano}

\authors{Ian W. McBrearty and Paul Segall}

\affiliation{1}{Department of Geophysics, 397 Panama Mall, Stanford University, Stanford, California 94305-2215, USA} \vspace{0.2 cm}

\correspondingauthor{I.W. McBrearty}{imcbrear@stanford.edu}

\begin{keypoints}
\item Forecasting Caldera Collapse
\item Deep Learning
\end{keypoints}

\begin{abstract}

During the three month long eruption of Kīlauea volcano, Hawaii in 2018, the pre-existing summit caldera collapsed in over 60 quasi-periodic failure events. The last 40 of these events, which generated Mw $>5$ very long period (VLP) earthquakes, had inter-event times between 0.8 - 2.2 days. These failure events offer a unique dataset for testing methods for predicting earthquake recurrence based on locally recorded GPS, tilt, and seismicity data. In this work, we train a deep learning graph neural network (GNN) to predict the time-to-failure of the caldera collapse events using only a fraction of the data recorded at the start of each cycle. We find that the GNN generalizes to unseen data and can predict the time-to-failure to within a few hours using only 0.5 days of data, substantially improving upon a null model based only on inter-event statistics. Predictions improve with increasing input data length, and are most accurate when using high-SNR tilt-meter data. Applying the trained GNN to synthetic data with different magma pressure decay times predicts failure at a nearly constant stress threshold, revealing that the GNN is sensing the underling physics of caldera collapse. These findings demonstrate the predictability of caldera collapse sequences under well monitored conditions, and highlight the potential of machine learning methods for forecasting real world catastrophic events with limited training data.

\end{abstract}

\section{Introduction}

A major goal in geophysics is to forecast hazardous events, including volcanic eruptions, landslides, and earthquakes.  The latter in particular has proven extremely challenging, and indeed, some have claimed it is impossible \citep{geller1997earthquakes}. The advent of machine learning (ML) techniques \citep{ren2020machinelearning} raises the question of whether such methods can find previously unidentified patterns or precursory signals that enable prediction of various geophysical phenomenon. For example, in recent years, ML methods have been used to predict the instantaneous stress and timing of laboratory earthquakes \citep{rouet2017machine, jasperson2021attention, shokouhi2021deep, laurenti2022deep, borate2023using}, the timing and size of simulated subduction zone earthquakes \citep{corbi2019machine, blank2021can}, as well as inferring the current eruptive state (erupting or not) of Piton de la Fournaise volcano from passive seismic data \citep{ren2020machine}.  A particular challenge for ML, and particularly deep learning (DL) methods applied to earthquakes and volcanic eruptions, is the paucity of repeated events usable as training data. This has limited the widespread applicability of ML methods for geophysical forecasting.

The 2018 Kīlauea, Hawaii eruption and caldera collapse offers a rich dataset for testing ML approaches for prediction. The extraordinarily high eruption rate led to the collapse of the volcano's summit in 62 discrete events over the course of three months, beginning in early May, 2018 \citep{neal20192018}.  Collapses of several meters were accompanied by Mw $> 5.0$ Very Long Period (VLP) earthquakes \citep{flinders2020VLP, wang2022VLP}, and coincided with inflationary (outward and upward) deformation at extra-caldera GPS and tilt-meter stations \citep{neal20192018, tepp2020, AndersonJohanson2022}.  Extensive Volcano Tectonic (VT) seismicity ($<$ M 4.5) ceased following collapses, but then gradually built up over the subsequent half day (Fig. 1), reaching a quasi-steady rate \citep{neal20192018}. Following collapse events, GPS and tilt stations exhibited exponentially decaying deflation (inward and downward) reflecting depressurization as magma flowed from the sub-caldera reservoir to the eruptive vent \citep{segall2021repeating, roman2021dynamics, AndersonJohanson2022} (Fig. 1). During the eruption, time varying seismicity rate and deformation data were used by the Hawaiian Volcano Observatory (HVO) staff to roughly anticipate collapse events in real time (Kyle Anderson person. comm). However, rigorous quantitative forecasts of the caldera collapse events were not attempted. Empirical failure forecast models of the 2018 Kīlauea collapses have been explored \textit{post hoc} in \citep{tepp2021}, however the authors concluded they were unable to forecast the timing of individual collapse events. \citet{fildes2022Nowcast} employed a `nowcasting' method for predicting the larger earthquakes at Kīlauea in 2018, yet concluded that this ``is not a useful hazard assessment technique'' for these events.

Here we assess the efficacy of DL for predicting the timing of collapse events at Kīlauea after training on previous collapse cycles, using a combination of GPS, tilt, and seismicity data. We focus on the last 39 caldera collapse events beginning on June 11, 2018, after the development of the ring fault surrounding the collapse block(s) was mainly complete. During this period inter-event times between collapse events ranged from 0.84 to 2.23 days \citep{neal20192018}, with a gradual lengthening of recurrence intervals towards the end of the eruption. We trained a graph neural network (GNN) to predict the time to failure, based on data from the first fraction of the cycle (e.g., 0.5 - 1 days), as well as testing predictions throughout the full cycle (e.g., up to 2.2 days). We train models using GPS, tilt, and seismicity data, and all datasets simultaneously, and evaluate the comparative performance. As an additional challenge of this dataset, we test (or validate) on the last 10 cycles, which primarily have longer durations (e.g., $>$ 1.4 days), while the training cycles primarily have shorter durations (e.g., $<$ 1.4 days). Our findings highlight that DL can effectively differentiate between short- and long-duration caldera collapse cycles during the 2018 Kīlauea eruption, both for training, and validation events. The DL-models outperform non-ML based forecasting methods that rely only on the prior distribution of inter-event times, and are also more effective than a Support Vector Machine (SVM) in generalizing to validation events. Additionally, we find that the extra-caldera tilt-meter is the most valuable dataset, and predictions steadily improve with increasing duration of input window times used.

While training neural networks on very limited number of datapoints is challenging, the topic is still of significant importance in geophysical forecasting due to the limited occurrence of repeating failure events on the same fault over the instrumental record. While the datasets may be limited, the use of DL in these cases is still of great importance, since the observed data often comes in high-dimensional form, with numerous different datatypes, noise sources, and signals of interest entangled together. While directly extracting usable summary information from such datasets through hand-crafted features is feasible in some cases, directly learning these relationships from data is more effective and scalable. DL further simplifies handling arbitrary dimensional inputs and outputs and prediction tasks. While the Kīlauea dataset of 39 caldera collapse events may represent a small dataset in typical ML settings, compared to elsewhere in the world, this is a large number of repeated earthquakes on a single fault with continuous well monitored conditions. Hence, this application is a promising target for exploring the potential of deep learning on limited data regimes in geophysical forecasting.

\section{Methods}

\subsection*{GNN Overview}

We use a Graph Neural Network (GNN) architecture \citep{battaglia2018relational} to predict caldera collapse times directly from the observed data. In our design, we subdivide an arbitrary time series into windows of overlapping features on the nodes of a temporal (linear) graph. Several layers of graph convolution are applied to mix information among the nodes (in a higher dimensional feature space), and then a global mean over the graph is taken, and a final layer predicts the cycle duration from this aggregate signal. Different input time series considered include GPS from five extra-caldera stations (Fig. 1), a single extra-caldera tiltmeter, and cumulative seismicity counts from the relocated catalog of \citet{ShellyThelen2019}. Our model takes as input any of these seven time series, including data from the start of the cycle to an arbitrary window length of $T$. We allow $T$ to vary from a minimum of 0.2 days to one hour before any of the collapse events. Since each input graph can represent a different length time series, each will typically have a different number of nodes. It is important to note, however, that the number of trainable weights of a GNN is independent of the size of the graph it is applied to, so the same GNN can be applied to arbitrary length input time series (or graphs of different number of nodes). This allows the GNN to be trained over a distribution of possible graphs (or time series lengths) and learn a single set of weights that is effective for any of the possible inputs.

GNNs are effective for the particular problem of failure time prediction by offering a few properties: (i). they can be trained with a small number of trainable weights, which promotes easier generalization for small datasets, (ii). they allow processing arbitrary length time series with no change to the internal parameterization, and are also generally more efficient than, e.g., RNNs, Transformers \citep{dwivedi2020generalization},  (iii). they exploit both local and global information, and distributed computing, to aggregate all of the data into a single prediction. In effect, the GNN is able to learn how to perform an integration over the entire time series, which allows incorporating all of the historical data information without explicitly imposing a weighting of how to treat old or new information (or observed data) differently. These traits make a GNN solution to this problem encouraging. In addition, since the prediction targets are smoothly varying in time (e.g., time-to-failure is a linearly decreasing function), the architecture is well suited for the task of predicting a gradually evolving time-to-failure, by simply gradually incorporating more observations as additional nodes on a linear graph, and updating its prediction. This structural feature encourages significant stability in the models outputs, such that predictions will only gradually vary over time, and not show wild fluctuations between adjacent time steps, as is common with other ML approaches.

We further stabilize the GNN by giving each node the additional feature of its elapsed time since the start of the cycle. This absolute time information allows the GNN to know how long it has already waited since the start of each cycle, and hence to learn the prior information of the distribution of possible recurrence times (from the training data) that are likely after a given lag time $T$. The GNN learns this prior information automatically, since whenever the last node of an input graph is at time step $T$, the GNN is only trained on cycles of duration $> T$ for this sample, which is a limited subset of all possible events, and hence have a mean duration conditional on $T$. This conditional likelihood of time-to-failure evolves throughout each cycle, and is effectively the `null model' if no data other than the current waiting time is supplied as input to the neural network. By including both this conditional likelihood feature, and observed data, our model more easily focuses on learning the `perturbation' around the conditional likelihood that's already available directly from inter-event statistics.

\subsection{Adjacency Matrices}

We construct a graph for each time-series by taking overlapping, moving windows of data (sampled at 1/60 Hz), to represent each node. We extract 30 minute windows, stepping forward by 5 minutes per node, resulting in graphs with $\sim$100's of nodes (e.g., $\sim$140 nodes per 0.5 day trace). Each of these `per-station', and `seismicity-rate' sets of nodes, which represent a window of data at a fixed time step as the node feature vector, are given edges linking adjacent nodes in time. Hence, the starting and ending nodes have degree one and all other nodes have degree two. To enhance the long-range flow of information, in addition to these temporal nearest-neighbor graphs, we also use cayley graphs, a type of `expander graph' \citep{deac2022expander}, at every other graph convolution layer. These graphs are sparse and regular (e.g., each node has four incoming and outgoing neighbors), but very well connected, and hence provide faster information transfer with negligible increased computational cost. The combination of these two graphs allows the model to understand both the continuous flow of information between adjacent nodes, as well as larger-scale structures and trends in the time series.

\subsection{Graph Convolution}

GNNs typically rely on message passing, or equivalently, graph convolution, to transfer information between connected nodes on the graph \citep{gilmer2017neural}. This technique simultaneously transforms the latent data on each node with learned fully-connected neural networks (FCN) while sharing and pooling information locally between connected nodes on the graph. This allows the GNN to learn how to extract salient features to share between nodes, while also aggregating information across the entire graph into a consensus signal suitable for the prediction problem \citep{battaglia2018relational}. When the model is trained on a distribution of variable input graphs, it also adapts these features to handle variations to the input graph structures.

Mathematically, we implement a convolution module similar to that used in the Spatial Aggregation module of \citet{mcbrearty2023earthquake}, which is a form of GraphSAGE \citep{hamilton2017inductive}. The update of the latent state of each nodes latent vector $\vect{h}_i^{(k)}$ at layer $k$ to layer $k + 1$ is given by

\begin{equation} \label{eq: graph convolution}
    \vect{h}_i^{(k + 1)} = \phi_{agg}(\vect{h}_i^{(k)}, \vect{t}_i, \text{MEAN}\{\phi_{msg}(\vect{h}^{(k)}_j, \vect{e}_{ij}, \vect{t}_j, \vect{z}) \text{ } \vert \text{ } j \in \mathcal{N}(i)\})
\end{equation}

\noindent where $\mathcal{N}(i)$ represents the neighborhood set for node $i$ (i.e., the set of nodes $i$ is linked to via the adjacency matrix), $\vect{e}_{ij}$ represents `edge data' between nodes $i$ and $j$, $\vect{t}_i$ is a temporal embedding vector, and $\vect{z}$ is a `global state' summarizing information from the entire graph at each layer. In \eqref{eq: graph convolution}, $\phi_{msg}$ and $\phi_{agg}$ are two FCN layers, of the form $\phi(\vect{h}) = \sigma(W\vect{h} + b)$, where $W$ and $b$ are the FCN weight matrix and bias vectors, and $\sigma$ is a Parametric Rectified Linear Unit (PReLU) activation function. The \text{MEAN} operation takes the mean over each neighborhood set, while preserving the feature dimension (e.g., \text{MEAN}: $\mathbb{R}^{|\mathcal{N}| \times F} \rightarrow \mathbb{R}^F$), so that a single `aggregate message' is received by each $i$th node from its neighborhood set. The `comma' operations in \eqref{eq: graph convolution} all represent concatenation across the feature dimension.

In our application, we define the edge data as $\vect{e}_{ij} = \vect{h}_i - \vect{h}_j$, which allows the difference in features between adjacent nodes (or time steps) to easily pass through the GNN layers, and $\vect{z} = \text{MEAN}_j \{\phi_{glb}(\vect{h}_j^{(k)}) \}$ is a global `summary' vector extracted from the whole graph at each layer, where $\phi_{glb}$ is a single-layer FCN, with a 5-dimensional output feature dimension. The temporal embedding, $\vect{t}_i$, is a 15-dimensional latent vector which is the output of a single two-layer FCN applied to the time step, $t_i$, of each node. This temporal embedding allows the GNN to be aware of what time step each node corresponds to in each input graph and incorporate this information in the message passing layers.

\subsection{Training the GNN}

We train the GNN using the Adam optimizer \citep{kingma2014adam}, with learning rate = 0.001, and updating for 1750 steps with a batch size of 50. In most cases, we consider the first 29 earthquake cycles for training, and the last 10 cycles for validation. For each sample of the batch during training, we sample a different cycle (from the first 29 cycles), and a different time series input length (discretized to 15 minute intervals). For each input length of $T$, only cycles with durations longer than $T + 1$ hour are included in the training data. 

In an additional test, we also train a series of models trained over a range of different fixed input length windows, and different number of training cycles, to compare the tradeoff of these parameters and the generalizablity of the model with respect to the amount of available training data. For these cases, since there is a respectively smaller dataset (compared to using arbitrary length input time windows), we update for 500 steps with a batch size of 15. The model is implemented in Pytorch Geometric \citep{fey2019fast}. Additional details on the number of layers and feature dimensions used in the GNN are listed in the Supplemental Materials. Due to the compact size of the model ($\sim$32,000 free parameters), and small number of training datapoints, the GNN takes $\sim$1.1 minutes to train on a single Tesla T4 GPU.

\section{Results}

\subsection{Prediction results with fixed window length inputs}

After training the GNN on the first 29 cycles, we evaluate its ability to predict the recurrence times of the remaining 10 collapse events with different fixed input length windows, considering either 0.5, 0.75, and 1.0 day windows, and using GPS and tiltmeter data as input (Fig. 2). Loss curves for this trained model are shown in (Fig. S1), which show that the model rapidly converges and can be stopped after $\sim$1500$ - $2000 update steps. We see effective performance of the GNN for all of the training data and reasonably accurate predictions for the validation data, with increasing accuracy with longer input data windows used. For input lengths of 0.5 and 1.0 days, the combined GPS and tilt model obtains average validation residuals of 3.89 hours and 1.75 hours, respectively (Table 1), with corresponding $R^2$ coefficient of determination values of 0.65 and 0.93 (Table S1). Notably, we have high accuracy on much of the validation data despite that due to the chosen data split between training and validation, much of the training data have significantly shorter recurrence times than the validation events. For example, other than the longest training event with duration 1.85 days, the rest of the training events have durations $< 1.45$ days, and eight of the 10 validation events have durations $>$1.45 days, with two of these having durations $>$2 days. Even for the validation events of $>$2 days (with durations of 2.23 and 2.16 days), with 1.0 day input length, we predict recurrence times of 2.18 and 2.03 days, respectively, and both values are accurate to within $\sim$1.2 and 3.3 hours (Fig. 2). These predictions are far in excess of the largest training cycles recurrence time of $\sim$1.8 days. When training with a random split between training and validation events that does not preserve chronological order, we see a $\sim$40$\%$ reduction in the validation residuals (Table S2), which highlights the challenge of handling the long-term drift of the cycles.

The GNN results are compared against a `null model' consisting of the mean of all prior recurrence intervals in Fig. 2d. The null model predictions are significantly less accurate compared to the GNN predictions. For example, for 0.5 day length inputs, the RMS residual of the null model on the validation data is 14.1 hours, compared to 3.9 hours with the GNN (Table 1). The null model struggles both in differentiating cycle length among the earlier training events, and handling the longer events towards the end of the eruption. Because the null model only relies on inter-event statistics of previous events, it is not effective for handling the gradual lengthening of cycles. In addition to this test, we also compare our results against a Support Vector Machine \citep{drucker1996support}, which takes the full time series up to time $T$ as input and directly predicts the duration through a kernel mechanism. This type of ML model is capable of non-linear regression on high-dimensional input data, and in theory, could also find the functional relationship between the input data and recurrence times. However, the SVM showed accurate training results (e.g., 1.51 hour residual) but poor validation results (e.g., 5.23 hour residual) with 1.0 day length inputs, indicating a challenge in generalizing between training and validation events (Fig. S5).

Results obtained using various combinations of GPS, cumulative seismicity, and tilt (Table 1) reveal that the best model results are obtained using both GPS and tilt data, while the model trained on only cumulative seismicity performs the worst. Notably, the model trained with seismicity counts in addition to GPS and tilt performs well on training, but worse than the GPS and tilt only model on validation data, indicting a possible distributional shift of the seismicity data towards the latter cycles of the eruption. The model trained on only tilt performs better than the GPS only model, which may be attributable to the high-SNR of the tilt data. The accuracy of the GNN trained on individual GPS stations is shown in Table S4, showing that the model trained with all GPS stations yields errors around $\sim$75\% of those obtained using only a single GPS station. Individually, stations UWEV and CRIM perform the best, while OUTL and AHUP perform the worst, revealing that station proximity to the caldera rim appears to correlate with accuracy (Fig. 1c).

\subsection{Prediction results with arbitrary window length inputs}

To assess the accuracy of the model at predicting the time-to-failure throughout the entire failure cycle, we plot the predictions as the input length increases from 0.2 to 2.2 days within the last 20 cycles (Fig. 3). At each time step we plot the estimated collapse time minus the current time, resulting in continuously updated estimates of `remaining time to failure' provided by the GNN. We observe that the GPS and tilt model has the most accurate and smoothly varying predictions on validation events compared to models trained with seismicity, GPS, or tilt data separately (Fig. 4). In addition, this model exhibits lower deviation of predictions between multiple repeated training runs compared to the other models, though some uncertainty remains on the longest two cycles of the validation events (Fig. 3).

Notably, we see that the seismicity-only model has very poor performance on validation events, and it appears the seismicity model has largely only learned the null model behavior (e.g., the null model has validation residual $\sim$14 hours and the seismicity model has validation residual $\sim$13.5 hours for 0.5 day inputs). Each predicted time-to-failure trace of the seismicity-only model is nearly the same, indicating the model is insensitive to the observed data and has primarily only learned the average conditional time-to-failure from the training events, explaining why it significantly under predicts the longer validation events (Fig. 4a). This insensitivity to cumulative seismicity counts may be attributable to the fact the seismicity counts are highly similar between each cycle and are largely linearly increasing in time with little variation between cycles (Fig. 1a). The model trained on a single GPS station, UWEV, is overall quite effective, and shows little deviation between repeated training runs of the model, though notably shows a strong transient change in its prediction for the fourth validation event in Fig. 4c that is not shown by the combined five GPS station model (Fig. 4b). This appears to correlate to a significant transient change in energy in the raw displacement time series observed on this station for this cycle (Fig. S3).

We additionally see that over elapsed time, the predictions tend to converge more closely to the ground truth cycle durations. To highlight this, rather than time-to-failure, we plot the estimated cycle duration directly output by the GNN at each time step during the validation cycles (Fig. 5). As can be seen, for all validation events, and even the two longest cycles $>$2 days, the GPS and tilt model predictions converge and often stabalize to within $\sim$2.5 hours of the true values after less than half of the cycle has elapsed. Many of the other models approach values close to the ground truth over time, but show less stability, and instead have a gradual increase in the output predictions with elapsed time, in some cases ultimately exceeding the target. To quantify the improvement in accuracy as collapse time approaches, we plot the average validation error as a function of time-before-failure (Fig. 6c), and the corresponding time-dependent $R^2$ coefficient of determination (Fig. 6d). This shows a clear decrease in residuals as failure time approaches, and a strong positive $R^2$ correlation between predictions and ground truth for all times within $\sim$1.2 days of collapse. The GPS and tilt model has $R^2 > 0.95$ in the last 24 hours before collapse, while the models trained with only GPS or tilt data separately have $R^2$ values varying between $\sim$0.8 and 0.93 over this interval. The GPS model gradually improves over time, while the tilt models performance slightly decays, indicating a possible challenge of generalizing between training and validation events when using only a single time series as input.

\subsection{Data Tradeoffs}

To explore the sensitivity of our model to the trade-off in the number of earthquake cycles used in training and the given temporal input window length used, we trained a model for a different pair of (\textit{number of cycles, time input length}), over a grid of values. We train the model five times for each pair, and report the average residual on training and withheld data (Fig. 6b). This test effectively reveals insight into the overfitting issue, as for some pairs (e.g., few cycles and long time windows) the model completely overfits and does not generalize to unseen data (e.g., bottom right of Fig. 6b); while for others, e.g., modest window lengths and number of training cycles, effective generalization does occur for held-out data (e.g., upper middle and right of Fig. 6b). Of particular note, including training cycle $25$ noticeably improves validation performance across all input window lengths, since this cycle is the longest of the training cycles at 1.85 days, while all other training cycles have duration $< 1.45$ days (Fig. 2). Since many of the validation events are also long cycles ($>$ 1.45 days), including this cycle in training appears important for helping the GNNs ability to estimate longer duration events. This tradeoff test further gives confidence that the models trained on the first 29 events and validated (or tested) on the later 10 have obtained true predictive capabilities and not only overfit the training data.

\section{Discussion}

\subsection{Predictability of Collapse}

The repeated caldera collapse events of the 2018 Kīlauea eruption offer an unprecedented dataset for testing forecasting methods with high precision local GPS, tilt, and seismicity data. Our analysis reveals that deep learning methods enable the prediction of caldera collapse times with high accuracy (e.g., Fig. 2). This result sheds new light on the possibility of real time forecasting of collapse events that was previously deemed out of reach (e.g., \citet{tepp2021, fildes2022Nowcast}). For example, our method estimates collapse times with $\sim$3.9 hour accuracy with only 0.5 days of data, or $\sim$1.8 hour accuracy with 1 day of data (Table 1). Within the last 24 hours before collapse, the mean validation residual is $<$ 1.5 hours (Fig. 6c). These errors correspond to $R^{2}$ coefficient of determination values of $>$0.95 (Fig. 6d). In contrast, the non-DL based SVM model has validation residuals of $\sim$5.5 and 5.2 hours for 0.5 and 1.0 day inputs, respectively, and shows signs of substantial overfitting to the training data (Fig. S5). Estimates based solely on the statistics of past events achieve $>$10 hour accuracy for these intervals, and have negative $R^2$ values.

A significant finding is that our model shows signs of extrapolation capabilities. For many of the validation cycles, the durations are $>$ 1.45 days, yet most training cycles have durations $<$ 1.45 days, and only one training cycle has a longer duration of 1.85 days. The two longest validation cycles are 2.16 and 2.23 days, and our model predicts 2.18 and 2.03 days, respectively, for these events (Fig. 2c). Hence, our model correctly predicts recurrence times for multiple collapse events with durations greater than the longest training cycle. This indicates a strong generalizability of the model to the prediction task of failure time at Kīlauea volcano, and not simply memorization of the training data distribution. An additional strength of our model is that it naturally predicts a smooth prediction of time-to-failure as more data (or longer graphs) are given as input, and can incorporate all of the historical data from the start of the cycle up until the current observed time. This sets it apart from previous laboratory earthquake prediction studies (e.g., \citet{rouet2017machine}), which used individual windows of data for each input, and as a result, had less smooth predictions in time. For a real-world forecast of time-to-failure, having smooth predictions in time makes the predictions more actionable, and using all of the observed data in the input rather than sliced windows of data can lead to a more careful understanding of the system dynamics. Our models approach to only gradually update its prediction as time elapses by attaching additional nodes to the input (e.g., Fig. S2) is a sensible approach for a stable time-to-failure prediction model.

An additional strength of our model is that the conditional prior distribution of inter-event times $>T$ for any input window of length $T$ is effectively built into the model through the training process. Since the prior distribution of all training cycle durations is information that is available during training, a real-world forecasting model should make use of this information. By incorporating the time step with each node (Eq. 1), the model can automatically learn, at a minimum, the conditional distribution of time-to-failure as a function of input length. This capability was made clear in the time-dependent predictions of the seismicity-only model (Fig. 4a), which were largely insensitive to the input seismicity count data, but still roughly predicted the average conditional likelihood of the training data for all validation cycles. For the other models, we can infer their predictions take into account the conditional likelihood of time-to-failure based on both the current waiting time, and the observed GPS and tilt data, and from this combination have achieved performance far in excess of using only the prior inter-event distribution.

\subsection{Forward Perspective}

The success of this study follows upon a recent trend of machine learning methods in geophysics enabling the prediction of a wide diversity of geophysical phenomenon \citep{rouet2017machine, corbi2019machine, ren2020machinelearning}. The potential for ML-enhanced geophysical forecasting methods is still accelerating, and accurate prediction of the timing of volcanic eruptions and earthquakes would be a remarkable breakthrough. This study takes one step further toward this goal by revealing the predictable behavior of the quasi-regular caldera collapse events at Kīlauea that generated Mw $> 5$ VLP earthquakes. This reveals that under optimal conditions, with enough observational data and suitably chosen models, catastrophic geophysical events, such as caldera collapse, are predictable in real time.

Other basaltic calderas have experienced episodic collapses, including Miyakejima, Japan \citep{munekane2016} and Piton de la Fournaise, Reunion Island \citep{Fontaine2014}. Similar trends in GPS and tilt data recorded during these events suggest that it should be possible to apply our GNN architecture to forecasting collapses at these and other volcanoes as well. In contrast, for tectonic faults we rarely have geodetic measurements of strain accumulation over multiple earthquake cycles of characteristic size, so training a model directly as used here for those settings will be challenging. However, it is possible that other approaches may be discovered. For example, training on synthetic data may be a viable approach (e.g., \citet{corbi2019machine, costantino2023seismic}), and training on an ensemble of fault systems may enable the GNN to identify common features that are indicative of system state (or likelihood of upcoming failure) that are not only specific to one site. In turn this could enable forecasts even on fault systems that do not have numerous rupture cycles recorded. 

In future applications, the approach could incorporate spatial-temporal graphs that link adjacent stations, and hence be generalizable to any number of stations, and flexible to the station geometry (e.g., \citet{zhang2022spatiotemporal, mcbrearty2023earthquake, sun2023phase, tan2023learning}). Such an approach could enhance the models ability to generalize to new sites without re-training, and also be adaptive to the incoming- and outgoing- of station data availability. During training, techniques such as domain adaptation \citep{farahani2021brief} and few shot learning \citep{wang2020generalizing} can be used to improve generalization between training events and unseen, future events, that may differ from training data due to non-stationary trends that are prevalent in volcanic and tectonic systems. Additionally, incorporating seismicity from enhanced seismic catalogs (e.g, \citet{retailleau2022automatic, wilding2023magmatic}) may prove valuable, and rather than using earthquake counts, it may be possible to use seismic point-clouds as an input to the GNN directly.

\subsection{Interpretation}

While we do not know the features the GNN uses to predict collapse times, we believe that the deformation data correlates with increases in stresses acting on the ring fault. Between collapse events magma flows from the sub-caldera reservoir to the eruption site, causing the pressure to decrease. This causes both deflationary deformation \citep{segall2019}, as well as an increase in the shear stress acting on the ring fault \citep{kumagai2001}. Collapse is believed to occur when a frictional threshold is reached \citep{segall2021repeating}. Hence, by this interpretation, we emphasize that we do not believe the GNN is detecting `precursory signals' in the sense the term is usually employed, but instead uses the GPS and tilt data as a proxy for tracking the change in driving stress acting on the fault. 

To test this, we generated synthetic GPS displacements consisting of exponential decays with additive noise. The amplitude of the signal was scaled to match observations at the various GPS sites, with characteristic decay times ranging from 0.2 to 1.1 days (comparable to the observed range), for which a corresponding stress history is known \citep{wang2022VLP}. The `noise' consisted of de-trended GPS displacements from a remote station with negligible signal. Our expectation was that the GNN trained on the real data would predict failure times corresponding to a constant stress threshold for these synthetic events, and this was largely the case (Fig. 6a). The predictions corresponded to a nearly constant stress threshold, which decreases slightly at the longest decay times, possibly because the model has less available training data from longer cycles. An unexpected result was that the stress threshold predicted was higher than anticipated based on fitting exponential decays to the GPS data for the same collapse cycles. A plot similar to Fig. 6a with exponential decays fit to the GPS data gives an average normalized stress level of 0.92 \citep{Segall2024stress}, which is less than in the synthetic test. This suggests that while the exponential decay of the GPS and tilt data is a key feature exploited by the GNN, there are other characteristics in the data aiding the forecasts.

\section{Conclusion}

Forecasting the timing of catastrophic geophysical events such as volcanic eruptions or earthquakes is a long standing challenge in geophysics. Recent applications of machine learning have shown promise in forecasting a wide range of geophysical phenomenon and has renewed interest in these endeavors. In this work, we demonstrated the potential of ML-based prediction of the timing of caldera collapse events during the 2018 Kīlauea eruption based on local GPS, tilt, and seismicity data. Our findings reveal that for this well monitored  sequence, cycle durations were (retroactively) predictable to within a few hours given data from only a fraction of the mean recurrence interval. The predictability of these large scale collapse events demonstrates the potential of ML-based forecasting of significant, real-world hazards. We expect that future application of these techniques on a broader array of geophysical forecasting problems will lead to increased insight, understanding, and predictability of complex systems.

\section{Acknowledgments}

We thank K. Anderson for discussion of the data and assistance in making Fig. 1, as well as helpful discussions with C. Johnson and T. Wang.

\newpage

\bibliography{bib}
\bibliographystyle{apa}

\newpage

\begin{table} 

\caption{Performance of the trained GNN for different combinations of training data, including GPS, tilt, and seismicity data. Models are trained five times and average residuals are reported for 0.5 days and 1.0 days of input data from the start of each cycle. The results of the SVM model and the null model based on the average duration of cycles longer than 0.5 or 1.0 days contained in the training data are also reported. Bold entries mark the best model in each column.}

\begin{minipage}{0.75\linewidth}

\vspace{0.5cm}

\begin{tabularx}{\textwidth}{l @{\hspace{2\tabcolsep}} c  @{\hspace{2\tabcolsep}} c  @{\hspace{2\tabcolsep}} c  @{\hspace{2\tabcolsep}} c  @{\hspace{2\tabcolsep}} c  @{\hspace{2\tabcolsep}} c  @{\hspace{2\tabcolsep}} c @{\hspace{2\tabcolsep}}}


\cmidrule(lr){1-6}

  & Residual Train. & Residual Vald. & & Residual Train. & Residual Vald. & \\
  & (RMS, Hours) & (RMS, Hours) & & (RMS, Hours) & (RMS, Hours) & \\
  & 0.5 Day Input & 0.5 Day Input & & 1.0 Day Input & 1.0 Day Input & \\
\cmidrule(lr){1-6}
 & & & & & & \\
Tilt & 1.36 ($\pm$ 0.18) & 4.44 ($\pm$ 0.62) & & 1.04 ($\pm$ 0.17) & 2.37 ($\pm$ 0.70) & \\
GPS & 1.27 ($\pm$ 0.17) & 5.46 ($\pm$ 0.76) & & 0.82 ($\pm$ 0.16) & 3.93 ($\pm$ 0.98) & \\
SVM & 1.88 ($\pm$0.00) & 5.57 ($\pm$0.00) & & 1.51 ($\pm$0.00) & 5.23 ($\pm$0.00) & \\
Seis. & 3.82 ($\pm$ 0.20) & 13.56 ($\pm$ 0.76) & & 3.63 ($\pm$ 0.19) & 11.82 ($\pm$ 0.93) & \\
Null & 5.05 ($\pm$0.00) & 14.09 ($\pm$0.00) & & 4.13 ($\pm$0.00) & 12.19 ($\pm$0.00) & \\
\cmidrule(lr){1-6}
& & & & & & \\
\bf{GPS, Tilt} & \bf{0.89} ($\pm$ 0.17) & \bf{3.89} ($\pm$ 0.76) & & \bf{0.57} ($\pm$ 0.20) & \bf{1.75} ($\pm$ 0.36) & \\
GPS, Tilt, Seis. & 1.06 ($\pm$ 0.33) & 5.18 ($\pm$ 0.80) & & 0.76 ($\pm$ 0.31) & 2.71 ($\pm$ 0.64) & \\

\cmidrule(lr){1-6}
\end{tabularx}

\end{minipage}
\end{table}

\newpage

\begin{figure}[ht!]
    \centering
    \includegraphics[width = 0.95\linewidth]{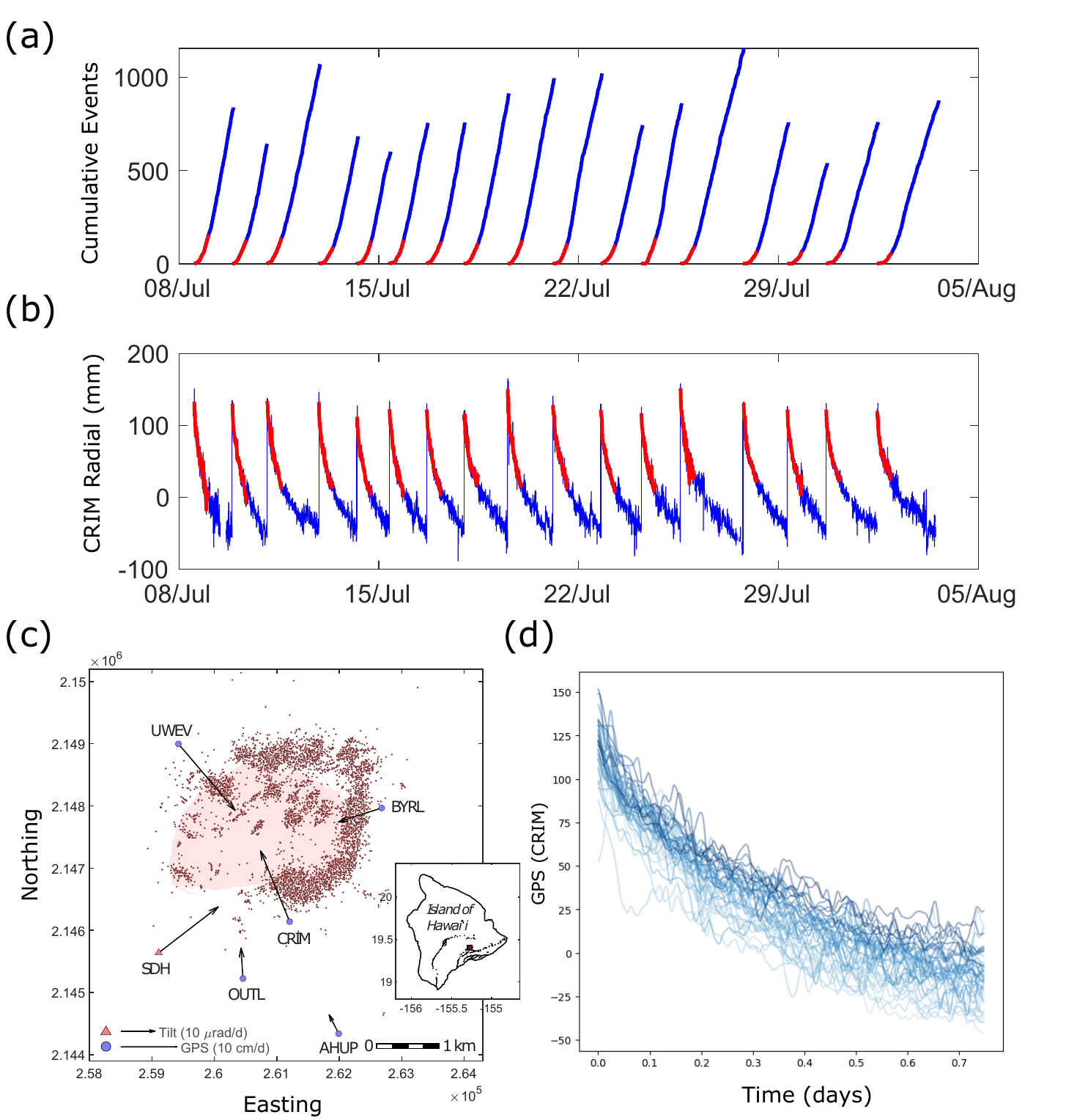}
    \caption{Cumulative seismicity (a) and radial displacement at GPS station CRIM (b) during the final 17 caldera collapse cycles of the 2018 Kilauea eruption. The red portion of each trace in (a,b) indicates an example 0.5 day input duration supplied to the GNN model. Map view of Kilauea along with the local seismicity  during the eruption (c), and the first 0.75 days of radial GPS time series from station CRIM (d), with line darkness proportional to the cycle length. For the GPS time series in (b), the red curves are smoothed by a moving 30 minute Gaussian filter for each collapse cycle (the filter does not combine data from adjacent cycles).}
    \label{Fig: Fig1}
\end{figure}

\begin{figure}[ht!]
    \centering
    \includegraphics[width = 0.95\linewidth]{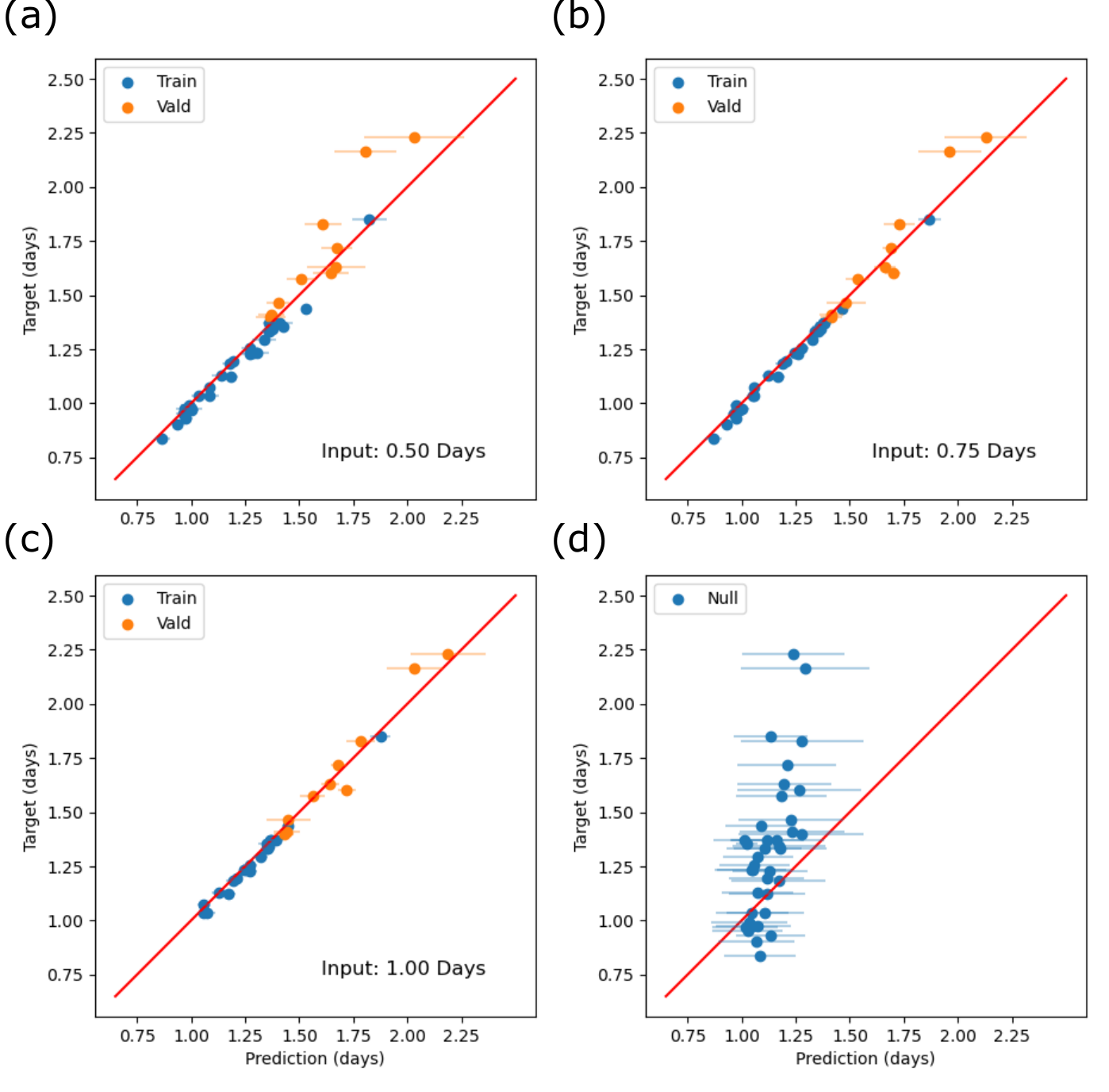}
    \caption{Comparison between predictions and observations for several different input window lengths. (a-c) GNN predictions for 0.5, 0.75, 1.0 day length inputs, respectively, for the GPS and tilt model. Predictions in (a-c) are average predictions from five repeated training runs of the model (with line widths marking $\pm$ 2 standard deviation). The first 29 events are used for training the GNN (blue), and the last 10 events are used for validation (orange). Results are compared with a `null' model (d), which for each cycle is the average duration of all previously observed cycles (with line widths marking $\pm$ 1 standard deviation).}
    \label{Fig: Fig21}
\end{figure}

\begin{figure}[ht!]
    \centering
    \includegraphics[width = 1.0\linewidth]{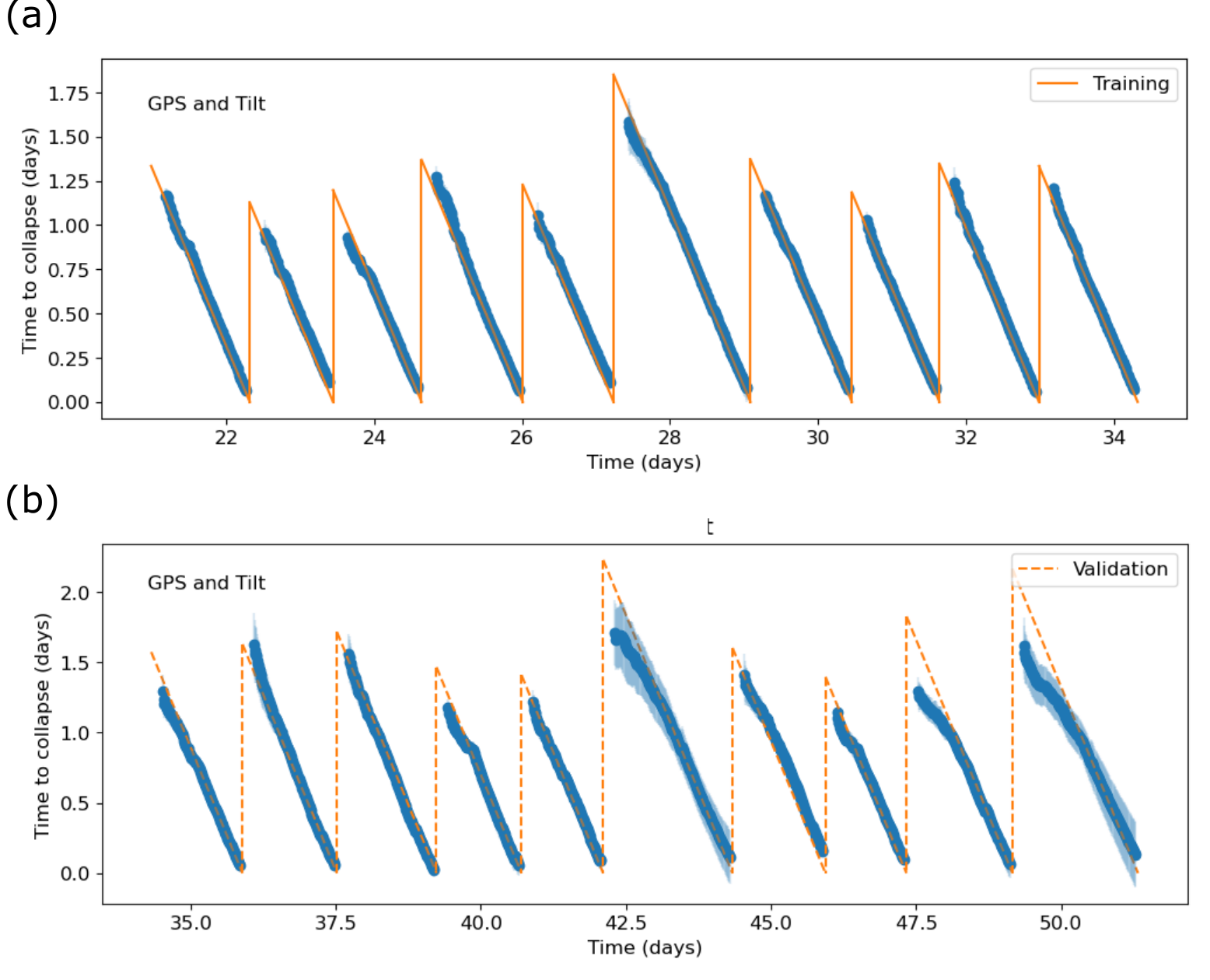}
    \caption{Predictions of the timing of the final 20 collapse events throughout each cycle using the GPS and tilt model, including 10 cycles from training and validation. At each time step, subtracting the current time from the GNN cycle duration prediction results in the updated time to failure (blue circles). The true time to failure is shown by orange lines (solid for training, dashed for validation). Each prediction has line widths marking $\pm$ 2 standard deviation based on five training runs of the model.}
    \label{Fig: Fig20}
\end{figure}

\begin{figure}[ht!]
    \centering
    \includegraphics[width = 1.0\linewidth]{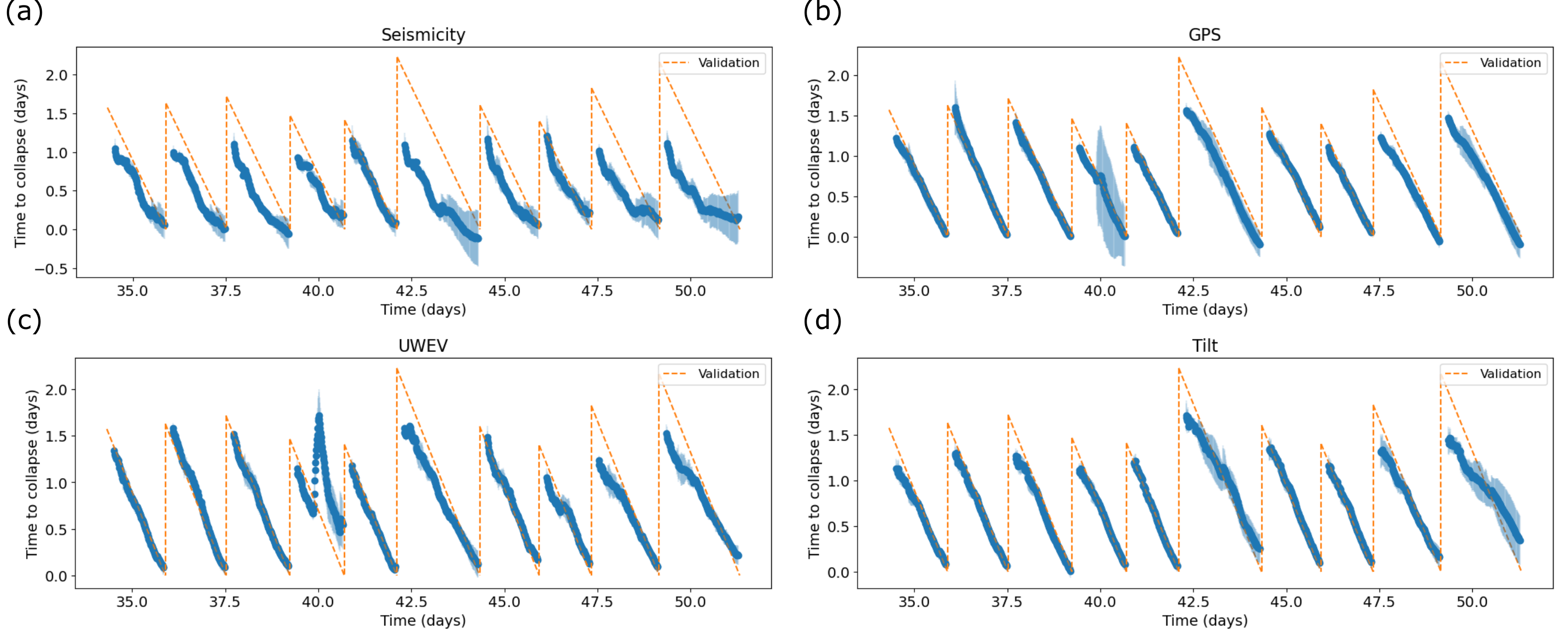}
    \caption{Predictions of the timing of the final 10 collapse events throughout each cycle (from validation) using different combinations of input data, including (a) seismicity, (b) all GPS stations, (c) GPS station UWEV, and (d) tilt. At each time step, subtracting the current time from the GNN cycle duration prediction results in the updated time to failure (blue circles). The true time to failure for these validation events is shown by dashed orange lines. Each prediction has line widths marking $\pm$ 2 standard deviation based on five training runs of the model.}
    \label{Fig: Fig20}
\end{figure}

\begin{figure}[ht!]
    \centering
    \includegraphics[width = 1.0\linewidth]{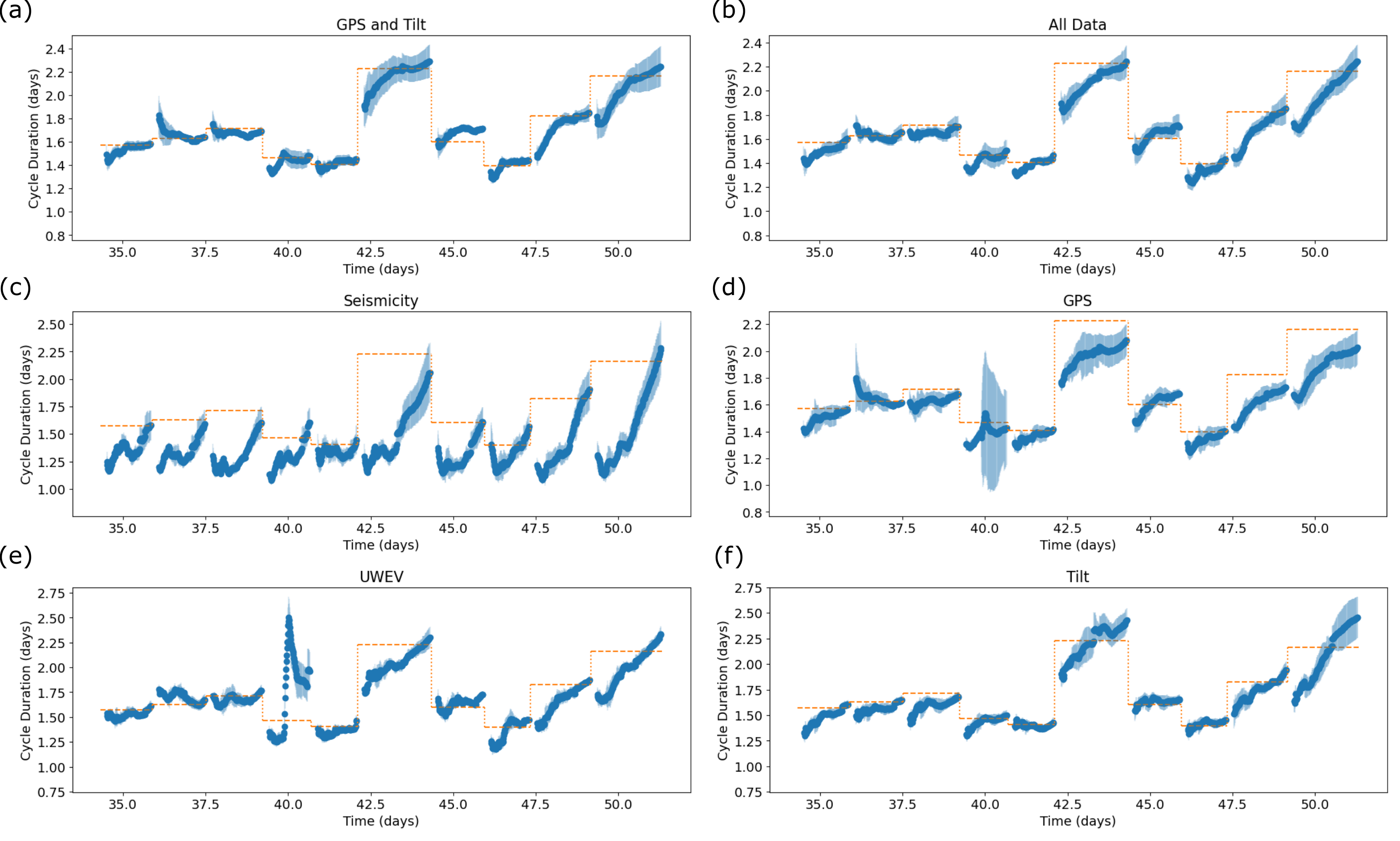}
    \caption{Predictions of the cycle duration of the final 10 collapse events (from validation) using different combinations of input data, including (a) GPS and tilt, (b) GPS, tilt and seismicity, (c) seismicity, (d) all GPS stations, (e) GPS station UWEV, and (f) tilt. At each time step, the absolute GNN cycle duration prediction values are shown (blue circles). The true cycle duration for these validation events is shown by dashed orange lines. Each prediction has line widths marking $\pm$ 1.5 standard deviation based on five training runs of the model.}
    \label{Fig: Fig20}
\end{figure}

\begin{figure}[ht!]
    \centering
    \includegraphics[width = 1.0\linewidth]{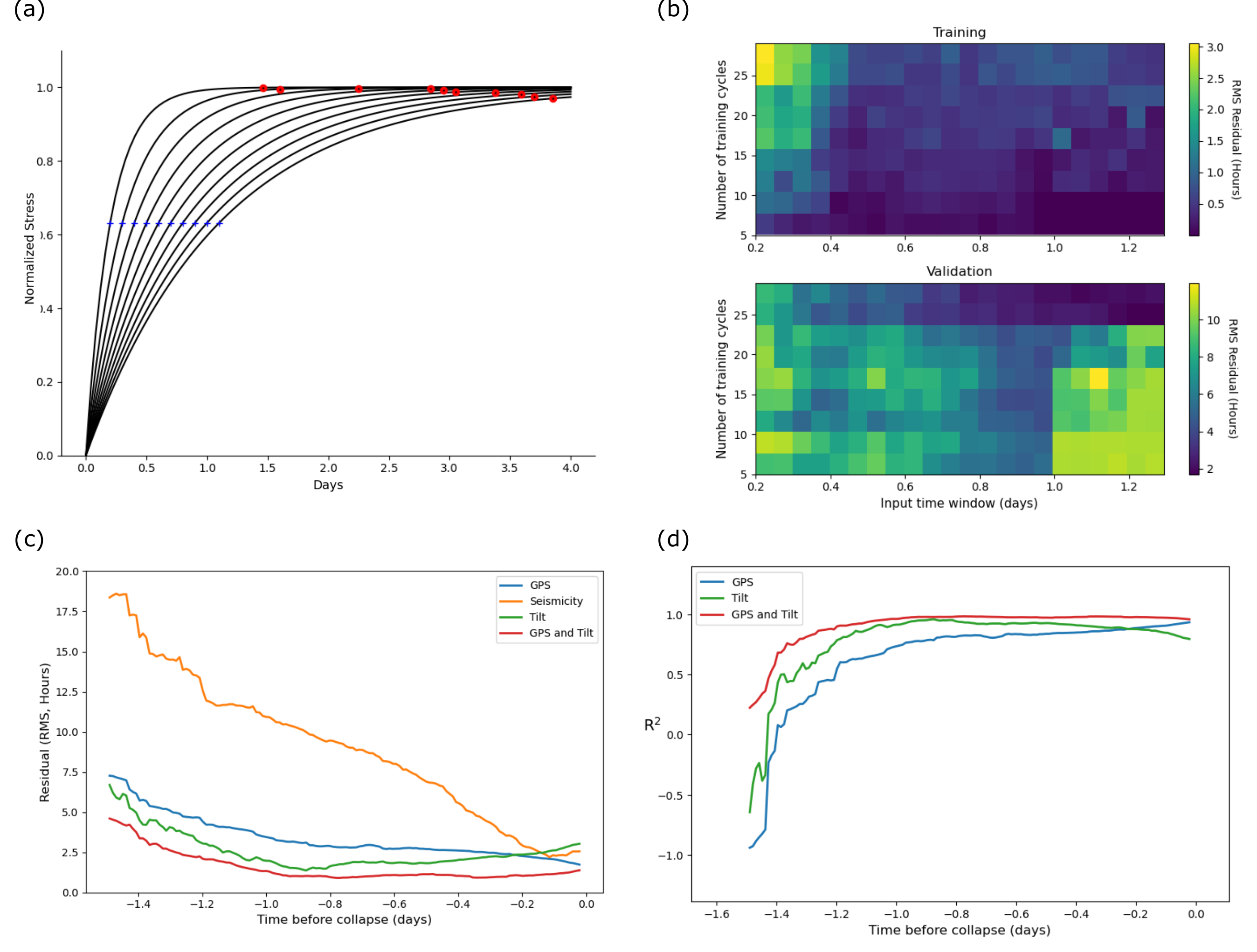} 
    \caption{Synthetic test and analysis of temporal-dependence of earthquake predictions. (a) Normalized stress as a function of time for synthetic exponential pressure decays, with characteristic decay times (blue $+$ symbols) and predicted failure times (red circles). (b). Trade-off in GPS and tilt model residuals on training and validation events, when training for only a specific combination of number of cycles and time window input lengths. (c). The average residuals on the validation events as a function of time-before-failure, and (d). the average coefficient of determination $R^2$ values on the validation events as a function of time-before-failure.}
    \label{Fig: FakeTest}
\end{figure}

\end{document}